\documentclass[prd,twocolumn,reprint,preprintnumbers,nofootinbib,floatfix]{revtex4-1}

\usepackage{subfigure,graphicx,amsmath,amssymb,hyperref}
\usepackage{youngtab}
\usepackage[utf8]{inputenc}
\usepackage{graphicx}

\usepackage{caption}
\usepackage{diagbox}
\usepackage{soul}

\usepackage[dvipsnames]{xcolor}

\usepackage{multirow}
\DeclareGraphicsExtensions{.pdf,.jpeg,.png}
\graphicspath{ {/images/} }

\newcommand{\bea}{\begin{eqnarray}}
\newcommand{\eea}{\end{eqnarray}}

\hypersetup{
    colorlinks=true,       
    linkcolor=blue,                 
    filecolor=blue,      
    urlcolor=blue,
    citecolor=blue
}

\begin{document}
\preprint{UCI-HEP-TR 2020-17}
\title{On Dark Matter Explanations of the Gamma-Ray  Excesses from the Galactic Center  and  M31
}

\author{Anne-Katherine Burns}
\email{annekatb@uci.edu}
\author{Max Fieg}
\email{mfieg@uci.edu}
\author{Arvind Rajaraman}
\email{arajaram@uci.edu}
\affiliation{Department of Physics and Astronomy,
University of California, Irvine, CA 92697-4575 USA
}

\author{Christopher M. Karwin}
\email{ckarwin@clemson.edu}
\affiliation{Department of Physics and Astronomy, Clemson University, Clemson, SC 29634-0978, USA}

\begin{abstract}
The presence of an excess 
$\gamma$-ray signal toward the Galactic center (GC) has now been well established, and is known as the GC excess. Leading explanations for the signal include mis-modeling of the Galactic diffuse emission along the line of sight, an unresolved population of millisecond pulsars, and/or the annihilation of dark matter (DM). 
Recently, evidence for another excess $\gamma$-ray signal has been reported toward the outer halo of M31. 
In this work we interpret the excess signals from both the GC and outer halo of M31 in the framework of DM annihilation, and show
that the two spectra  are consistent with a DM origin once $J$-factors are taken into account. 
We further compare the excesses to 
models of DM annihilation, and determine the corresponding best-fit parameters.
We find good fits to the spectrum both in two body and four body annihilation modes.
\end{abstract}

\maketitle

\section{Introduction}

There is now overwhelming evidence that most of the matter in the Universe is composed of dark matter (DM) ~\cite{Riess:1998cb,perlmutter1999measurements,Clowe:2006eq,Hinshaw:2012aka,Ade:2015xua,Eisenstein:2005su,Anderson:2013zyy,PhysRevD.98.030001,Peacock:2001gs}. However, despite much experimental effort for many decades now, the nature of DM
 still remains elusive. Determining the characteristics of DM  is one of the most important outstanding problems in particle physics. 

One of the most promising approaches for detecting DM is indirect detection. DM particles can annihilate or decay to Standard Model (SM) particles, which can be detected in astrophysical searches. In particular, annihilation or decay into photons gives a striking signal, since their direction of arrival is correlated with their annihilation location, and because photons can travel over large distances.

Simulations predict that the highest DM density should be near the Galactic center (GC), though models differ on the exact profile shape. Since the annihilation signal goes as the square of the density, the GC is thus expected to be one of the brightest sources of $\gamma$-rays from DM annihilation, and this makes it is an important target for indirect searches.

It has now been well established that there exists an excess of $\gamma$-rays toward the GC (as compared to the expected background)~\cite{2009arXiv0910.2998G, 
2011PhLB..697..412H,
2011PhRvD..84l3005H,
2012PhRvD..86h3511A,
2013PDU.....2..118H,
2013PhRvD..88h3521G,
2013arXiv1307.6862H,
2016PDU....12....1D,
2014PhRvD..90b3526A,
TheFermi-LAT:2015kwa,
2015PhRvD..91l3010Z,
2015JCAP...03..038C,
2015JCAP...07..013A,
2015PhRvD..91f3003C,
2016PhRvD..94f3504C,
Das_2017}. Intriguingly, the signal is found to be broadly consistent with having a DM origin, in regards to the energy spectrum and morphology. However, there are other plausible interpretations of the excess, including mis-modeling of the foreground emission from the Milky Way (MW), and an un-resolved population of point sources, such as millisecond pulsars~\cite{List:2020mzd,2019ApJ...871L...8F,2018MNRAS.475.5313F}. These other possibilities make it very difficult to extract a DM signal with a high degree of confidence.

Determining whether or not the GC excess does in fact have a DM origin (at least in part) will likely require complementarity with other targets, as well as other search methods (e.g.~direct detection). For $\gamma$-ray searches, the MW's dwarf spheroidal (dSph) satellite galaxies offer another promising target, as they are expected to be dominated by DM, with very little astrophysical background. However, thus far there has been no global signal detected, a result that is in tension with the DM interpretation of the GC excess ~\cite{Ackermann:2015zua,Fermi-LAT:2016uux}. However, it is important to note that the limits from the dSphs are subject to systematic uncertainties relating to their DM content, and this prohibits their ability to robustly constrain the GC excess~\cite{2020arXiv200211956A,Bonnivard:2014kza,PhysRevD.95.123012}.  

Looking beyond the MW, the Andromeda galaxy (also known as M31) is the closest large spiral galaxy to us and is predicted to be the brightest extragalactic source of DM annihilation~\cite{lisanti2018search,lisanti2018mapping}.   Recently, observations towards M31's outer halo reported evidence for an excess signal, with a peak in the $\gamma$-ray spectrum at an energy similar to the GC excess~\cite{2019ApJ...880...95K}. Moreover, the analysis is based on the outer regions of M31 where backgrounds from standard astrophysical emission are less dominant. It is thus plausible that both these signals result from DM annihilation.

In this work we perform a simultaneous analysis of the GC and M31 to determine if these two excesses are consistent with having a DM origin. 
We first examine the two spectra and see if they are consistent with each other once $J$-factors are taken into account. This turns out to be the case; furthermore, the required scaling turns out to be within the allowed range from a recent analysis of the M31 $J$-factor~\cite{Karwin:2020tjw}.

We then compare these spectra to various models of DM annihilation, We first consider 2-body final states, such as DM annihilating to bottoms and taus.
It is also interesting to consider four-body final states, which are motivated in models where DM which is coupled to the SM through
pseudoscalar mediators
\cite{karwin2017dark, escudero2017updated, Abdullah:2014lla, Rajaraman:2015xka} (such models avoid direct detection constraints).
We therefore
consider a few motivated examples of  annihilation to four final-state particles.
As we shall show, both two-body annihilations 
and four-body annihilations can produce good fits to the observed spectra.

The paper is organized as follows. In  section~\ref{sec:data_selection} we
review the observational data leading to the GC excess, and the more recent signal towards the outer halo of M31. We also review the current bounds on the $J$-factors of the two signal regions.
In the following section,
Section~\ref{sec:Method},
we
compare the M31 and GC spectra and examine whether they are consistent with the allowed $J$-factors. We then, in
Section~\ref{sec:dark_matter_models},
consider 
the spectra from specific DM models, for example, annihilation to b-quarks, and find the best fit to the observations. We end with a summary of our results. In Appendix A, we present results for other two-body and four-body annihilation channels, along with a table summarizing all our results.

\section{Review of GC and M31 
Observations} 
\label{sec:data_selection}
\subsection{GC}
For the GC excess we use data from Ref.~\cite{TheFermi-LAT:2015kwa}. Here we summarize a few main aspects of the analysis. The observations are based on approximately 5.2 years of \textit{Fermi}-LAT data, with energies between 1--100 GeV, in 20 logarithmically spaced energy bins. 

A majority of the diffuse emission in the Galaxy is due to the interaction of cosmic rays (CRs) with the interstellar gas and radiation fields. Indeed, the emission toward the GC is dominated by standard astrophysical processes, and the GC excess only amounts to a small fraction of the total emission. To quantify the uncertainty in the foreground/background emission, Ref.~\cite{TheFermi-LAT:2015kwa} employs the CR propagation code GALPROP\footnote{Available at \url{https://galprop.stanford.edu}}~\cite{gal1,gal2,gal3,gal4,gal5,gal6,gal7,gal8,gal9,gal10,gal11} to build four different interstellar emission models (IEMs), corresponding to two main systematic variations.

First, Galactic CRs are thought to be accelerated primarily from supernova remnants (SNRs) via diffusive shock acceleration {(see Ref.~\cite{TheFermi-LAT:2015kwa} and references therein).}
However, the distribution of SNRs is not well determined due to the observational bias and the limited lifetime of their shells, and so other tracers are often employed. Ref.~\cite{TheFermi-LAT:2015kwa} uses
two possible tracers, namely, the distribution of OB stars, which are progenitors of supernovae, and pulsars, which are the end states of supernovae. 

The second main variation comes from the tuning of the IEMs to the $\gamma$-ray data. This was done outside of the signal region, working from the outer Galaxy inward, with two variations in the fit. In the intensity-scaled variation, only the normalizations of the IEM components were left free to vary. In the index-scaled variation, additional degrees of freedom were given to the gas-related components interior to the Solar circle 
by also freely scaling the spectral index. 

\begin{figure}[t]
    \centering
    \includegraphics[width=0.5\textwidth]{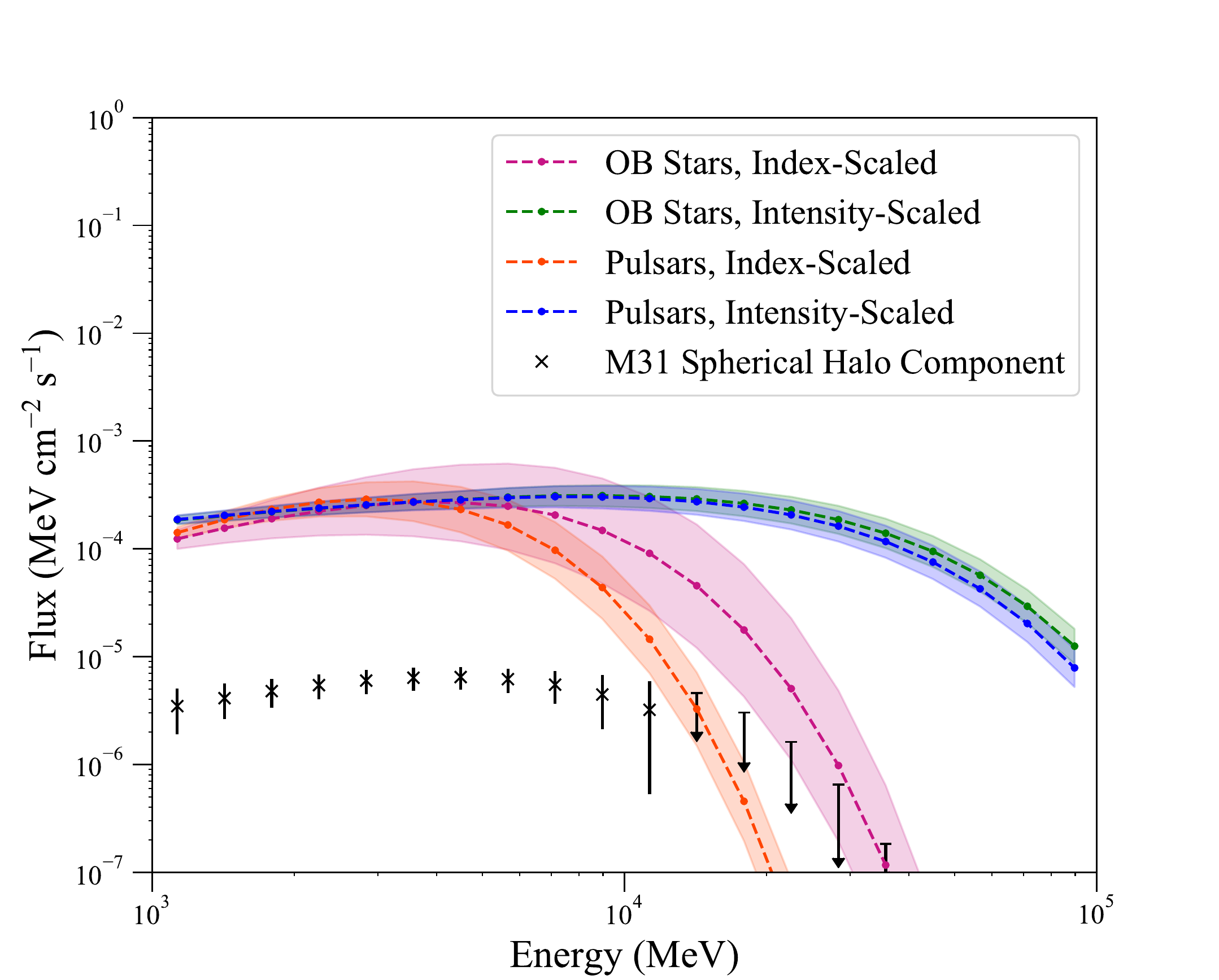}
    \caption{The colored dashed lines show the spectra of the GC excess for different IEMs, based on Ref.~\cite{TheFermi-LAT:2015kwa}. The bands show the 1$\sigma$ uncertainty. Black points show the  spectrum for M31's spherical halo component, based on Ref.~\cite{2019ApJ...880...95K}.
   }
    \label{fig:M31_vs_MW_unadjusted}
\end{figure}

Combining the variations in the CR density and the tuning procedure gives  four possible IEMs which quantify the uncertainty in the foreground/background emission. We shall denote these four IEMs as
(a) OB Stars, index-scaled (b) OB Stars, intensity-scaled (c) Pulsars, index-scaled; and (d) Pulsars, intensity-scaled. Figure~\ref{fig:M31_vs_MW_unadjusted} shows the spectra for the GC excess, corresponding to the four IEMs. Note that the intensity-scaled models have a high energy tail which is not present in the index-scaled models.

\subsection{M31}

For the M31 analysis we follow Ref.~\cite{2019ApJ...880...95K}. The analysis employs 7.6 years of \textit{Fermi}-LAT data, with energies between 1--100 GeV, in 20 logarithmically spaced energy bins. Similar to the GC,
the foreground emission from the MW is the dominant component when looking towards M31's outer halo, and Ref.~\cite{2019ApJ...880...95K} again used GALPROP to build specialized IEMs to characterize the emission.

Evidence for an excess signal was found, having a radial extension  of  $\sim 120-200$ kpc from the center of M31. 
To characterize the excess, three additional signal
components were added to the model (i.e.~in addition to the IEM). 
For the inner galaxy a $0.4^\circ$ disk was used, consistent with what has previously been reported~\cite{Fermi-LAT:2010kib,Pshirkov:2016qhu,Ackermann:2017nya}. A second concentric ring was also added, extending from $0.4^\circ$ to $8.5^\circ$ (corresponding to a projected radius of $\sim$120 kpc); this is referred to as the spherical halo component. Finally, a third concentric ring was added, extending from $8.5^\circ$ and covering the remaining extent of the field (corresponding to a projected radius of $\sim$200 kpc); this is referred to as the far outer halo component. 

Here, we only use data from the spherical halo region, and the corresponding spectrum is shown in Figure~\ref{fig:M31_vs_MW_unadjusted}. The inner galaxy is problematic for two reasons. First, it is difficult to disentangle a possible DM signal from standard astrophysical emission. Secondly, there still remains a high systematic uncertainty to the actual $\gamma$-ray signal that is detected~\cite{2019ApJ...880...95K,Karwin:2020tjw}.
This is due to an uncertainty in the underlying H I gas maps that are used for the Milky Way (MW) foreground. 
We also ignore the far outer halo region because it begins to approach the MW disk toward the top of the field, which significantly complicates the analysis. If the excess $\gamma$-ray emission observed toward M31's outer halo does in fact have a physical association with the M31 system, then it is particularly important to establish this in the spherical halo region~\cite{Karwin:2020tjw}. 

\subsection{J-factors}
\label{J-facts}
The greatest uncertainty for the DM interpretation of M31's outer halo comes from the $J$-factor. This is covered in extensive detail in Ref.~\cite{Karwin:2020tjw}. Here, we use results from that study to quantify the full uncertainty range, and below we summarize the key points. 

The $J$-factor characterizes the spatial distribution of the DM, and is given by the integral of the mass density squared, over the line of sight. When describing the DM distribution as an ensemble of disjoint DM halos,  
the $J$-factor is:
\begin{equation} \label{eq3}
J=\sum_i\int_{\Delta\Omega}d\Omega\int_{\text{LoS}}ds\rho_i^2(\mathbf{r}_i(s,\mathbf{n})),
\end{equation}
summed over all halos in the line of sight (LoS), where $\rho_i(\mathbf{r})$ is the density distribution of halo $i$, and $\mathbf{r}_i(s,\mathbf{n})$ is the position within that halo at LoS direction $\mathbf{n}$ and LoS distance $s$. 

$J$-factors determined from these spherically-averaged profiles are an underestimate of the total $J$-factor because of the effect of the non-spherical structure. This underestimate is typically encoded with a boost factor. The substructure component is very important for indirect detection, as it enhances the overall signal, since the predicted $\gamma$-ray flux scales as the mass density squared. This is especially true for MW-sized halos and toward the outer regions. 

The main uncertainties in the boost factor include the minimum subhalo mass, the subhalo mass function, the concentration-mass relation, the distribution of the subhalos in the main halo, the mass distribution of the subhalos themselves, and the number of substructure levels. In Ref.~\cite{Karwin:2020tjw} these physical parameters are varied within physically motivated ranges (as representative of the current uncertainty found in the literature) in order to quantify the uncertainty in the substructure boost. Additionally, there is also an uncertainty in the halo geometry, which is quantified 
by calculating $J$-factors for the different experimental estimates found in the literature.

In addition to the substructure and halo geometry, another primary driver of the $J$-factor uncertainty for obervations toward M31's outer halo is the contribution to the signal from the MW's DM halo along the line of sight, which is also accounted for in Ref.~\cite{Karwin:2020tjw}. Including all these uncertainties, the $J$-factor integrated over the spherical halo region, 
{(which we will henceforth denote as $J_{M31}$)}
is found to range from $(2.0-31.1) \times 10^{20} \mathrm{\ GeV^2 \ cm^{-5}}$, with a geometric mean of $7.9 \times 10^{20} \mathrm{\ GeV^2 \ cm^{-5}}$. 
We emphasize that this range accounts for the contribution from the MW's halo along the line of sight.

For the GC we use the $J$-factor from Ref.~\cite{TheFermi-LAT:2015kwa} that was used to extract the excess signal (which is consistent with the data that we use in this analysis). This corresponds to an NFW density profile with a slope $\gamma=1$, a scale radius $r_\mathrm{s}=20 \ \mathrm{kpc}$,  and local DM density $\rho_\odot = 0.3 \ \mathrm{GeV \ cm^{-3}}$. The $J$-factor integrated over the $15^\circ \times 15^\circ$ GC region
{ (which we will henceforth denote as $J_{GC}$)}
has a value of $2.2 \times 10^{22} \mathrm{\ GeV^2 \ cm^{-5}}$. 
We note that there is an uncertainty in the GC $J$-factor due to the value of the local DM density, as well as the other parameters in the density profile. However, in this work we consider just the uncertainty in the M31 $J$-factor, since it is dominant.

As described in more detail in the next section, a particularly important quantity in our analysis will be the ratio of the $J$-factors: 
\bea
J_r\equiv J_{GC}/J_{M31}
\eea
For the values of  
$J_{M31}$
between $(2.0-31.1) \times 10^{20} \mathrm{\ GeV^2 \ cm^{-5}}$, and
$J_{GC}=2.2 \times 10^{22} \mathrm{\ GeV^2 \ cm^{-5}}$
we 
find that 
$J_r$ lies between $J_{r,low} = 7.07$
and $J_{r,high} = 110.0$. Using the geometric mean of $J_{M31} = 7.9 \times 10^{20} \mathrm{\ GeV^2 \ cm^{-5}}$, we define $J_{r,mid} = 28.17$.

\section{Spectral Comparison of the GC and M31 Excesses}
\label{sec:Method}
\subsection{Best-fit J-factor ratios}
\label{Model Independent Scaling Factor Determination}
 
\begin{figure*}
  \centering
  \includegraphics[width=.4\linewidth, height=4cm]{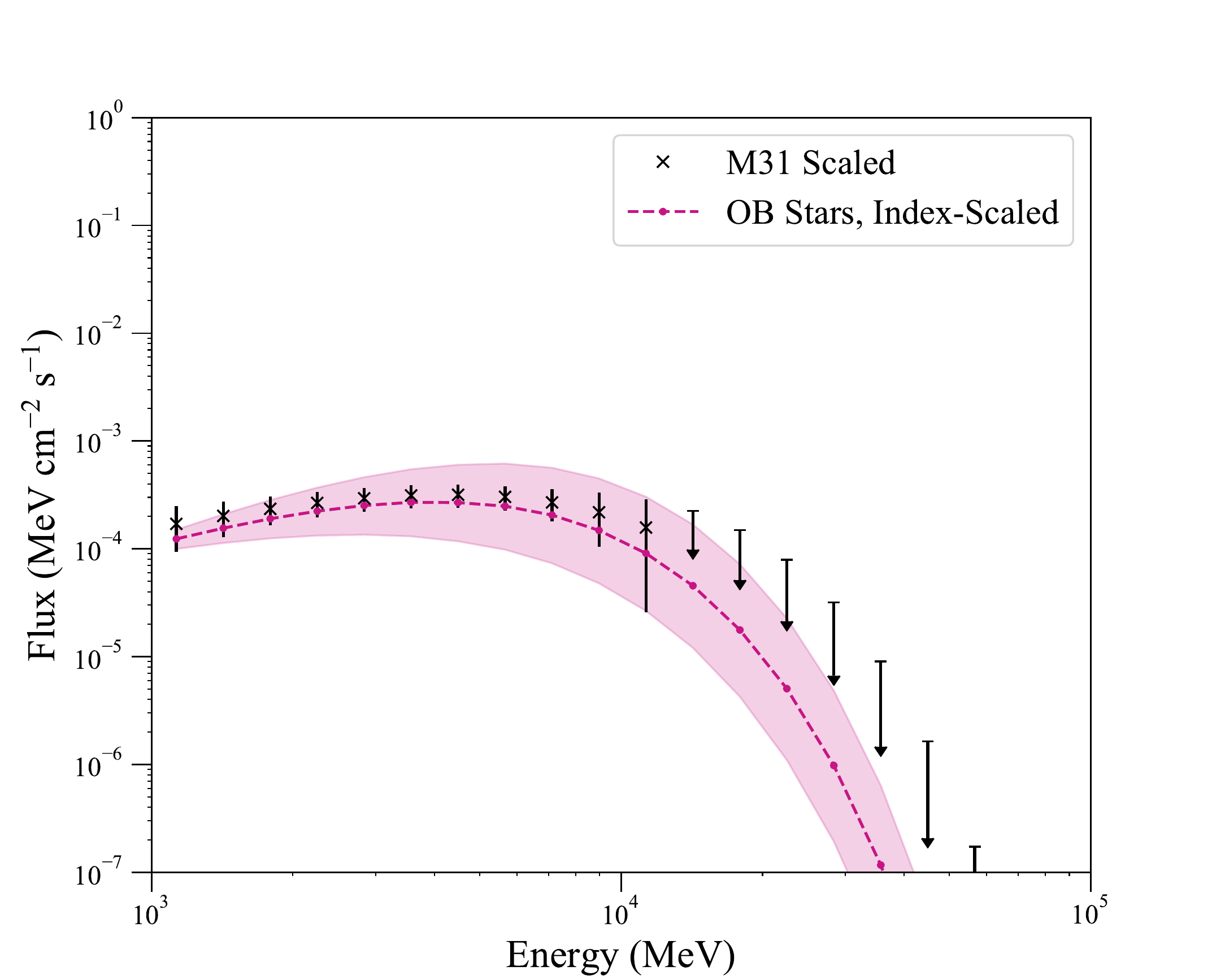}
    \includegraphics[width=.4\linewidth, height=4cm]{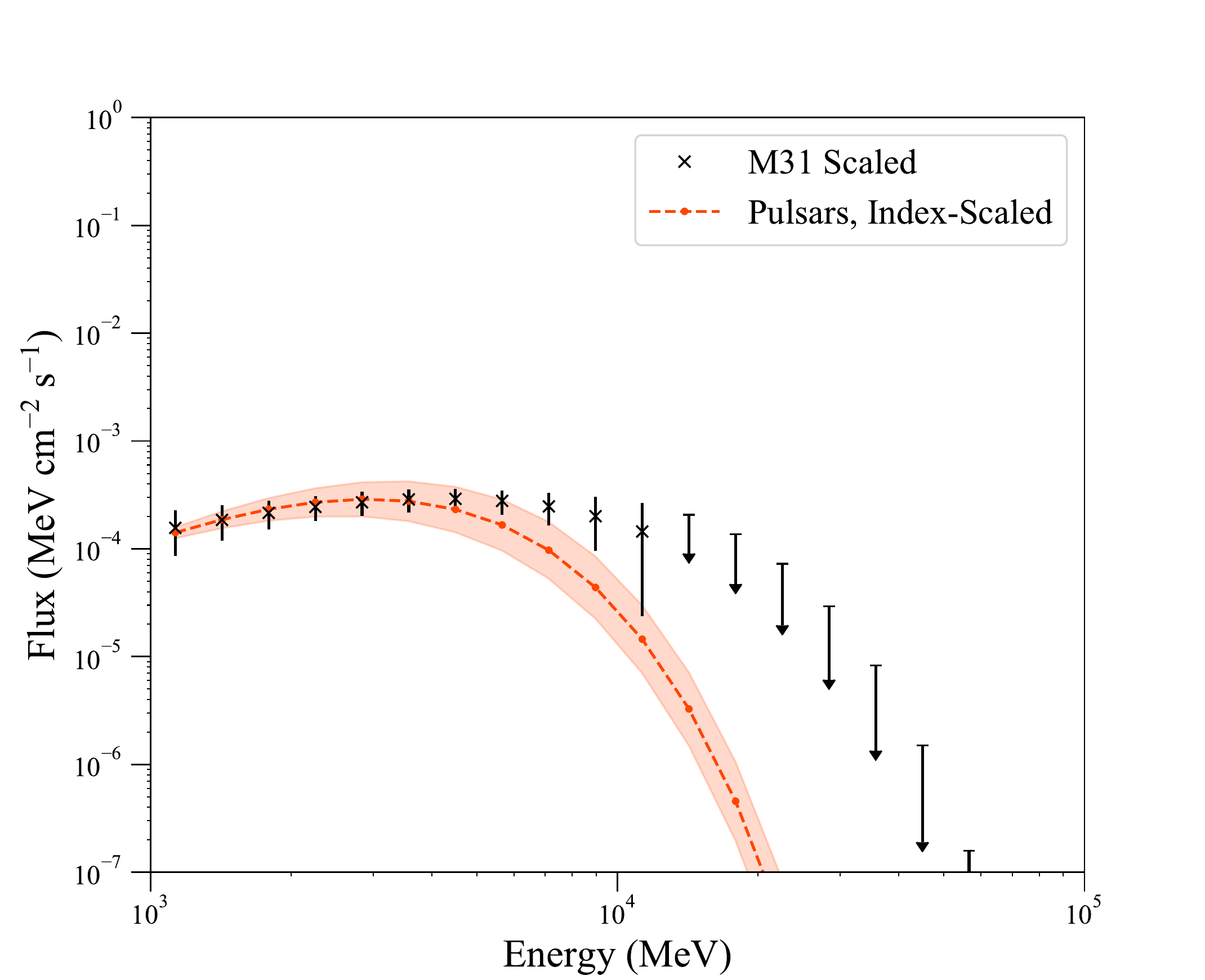}
  \centering
  \includegraphics[width=.4\linewidth, height=4cm]{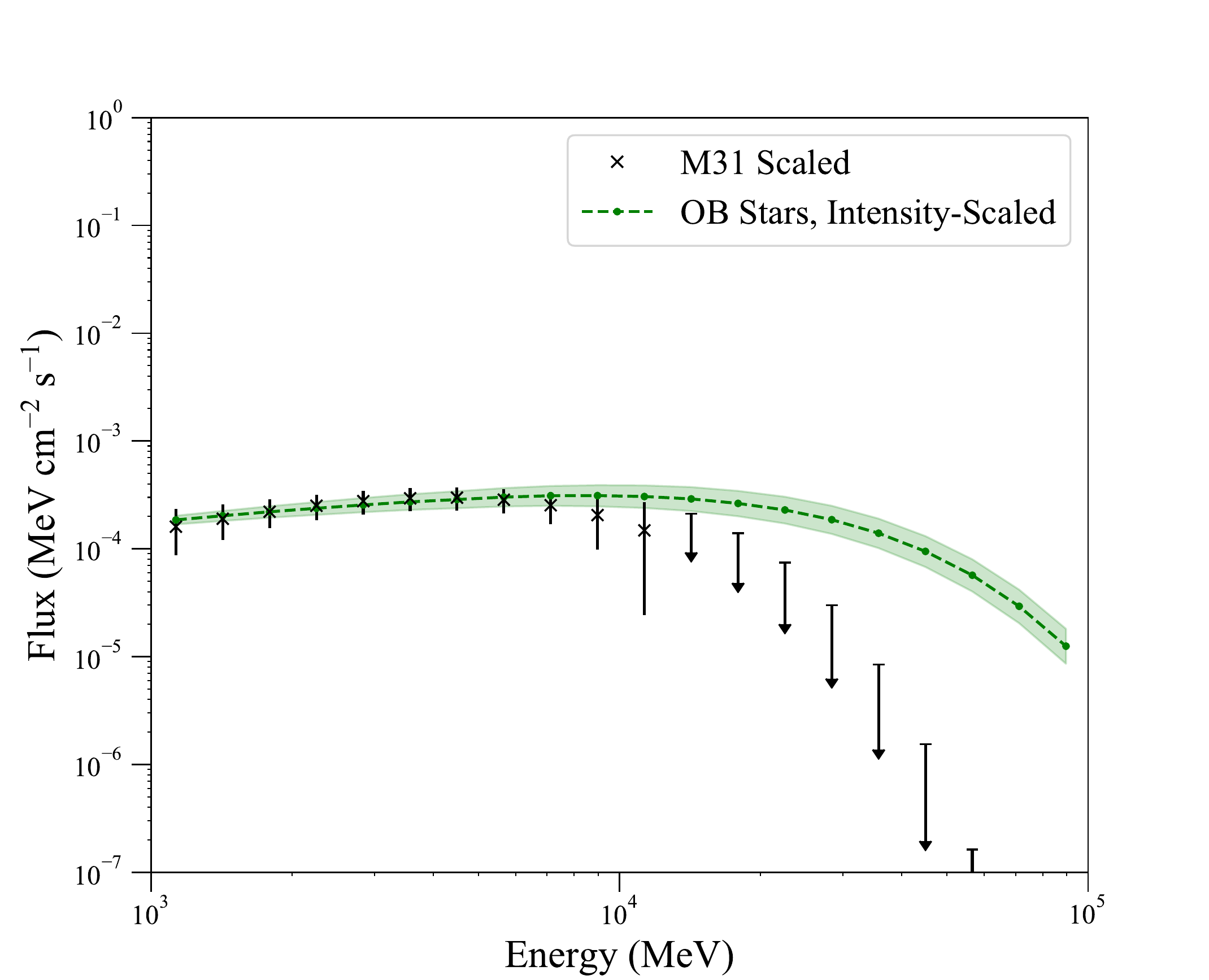}
    \includegraphics[width=.4\linewidth, height=4cm]{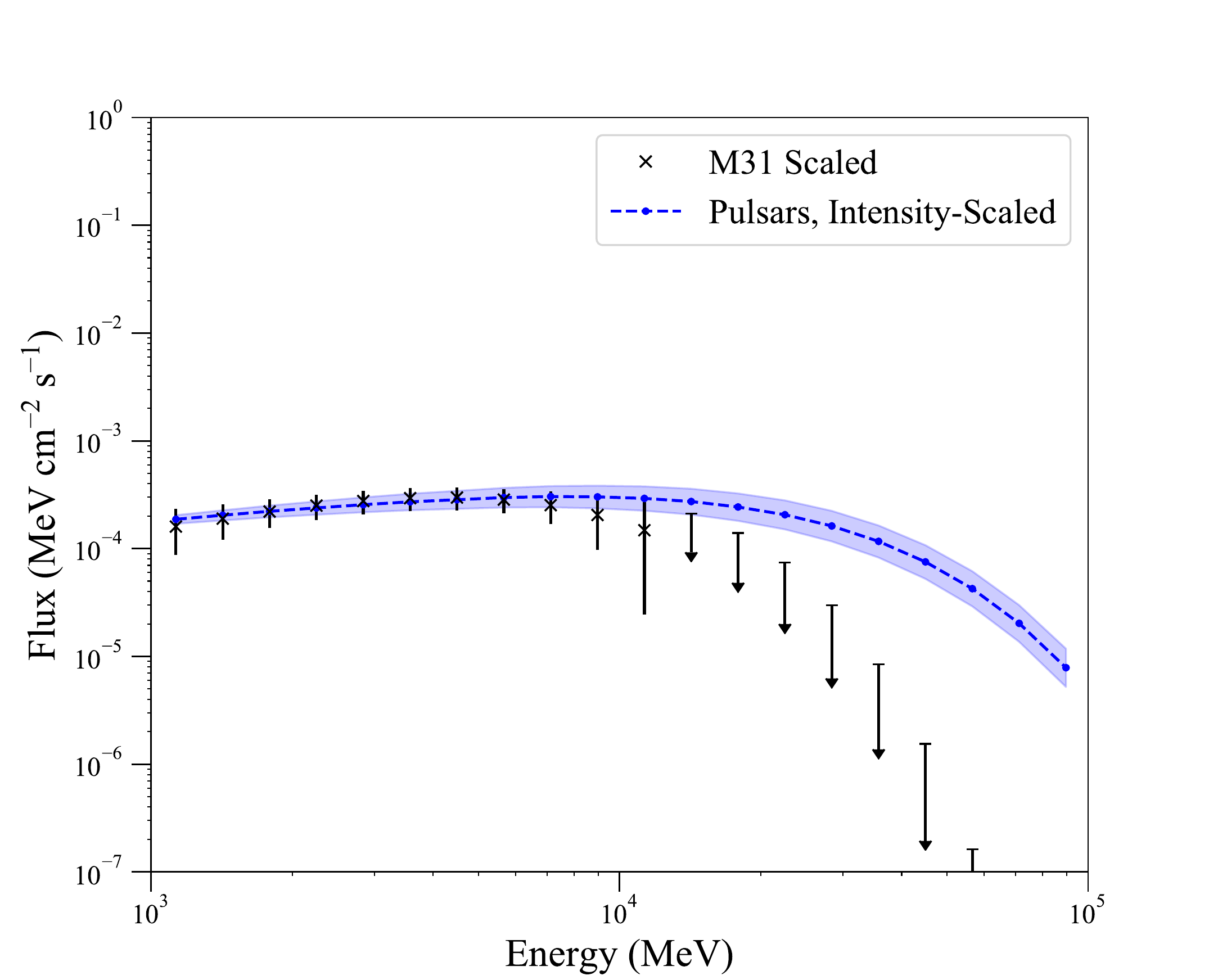}
  \caption{Comparison of GC spectra (colored bands) to scaled M31 spectra (black data points). The bands and error bars give the $1\sigma$ statistical error. The top two panels show the index-scaled IEMs, and the bottom two panels show the intensity-scaled IEMs. In each case the M31 data is scaled by the appropriate $J$-factor
  for the IEM.
  }
\label{fig:intensityvsM31}
\end{figure*}

The flux observed from M31 is much lower than that of the GC excess.
 If the excesses are indeed from an underlying DM model, then the underlying cross-section for DM annihilation to photons should
 be the same. The difference in the spectra 
 would then be attributable mostly to the ratio between $J_{M31}$ and $J_{GC}$. We note, however, that there may be some differences that arise from secondary emission, which depends on the particular astrophysical backgrounds in each respective 
 targets (i.e.~the gas and interstellar radiation fields)~\cite{Cirelli:2013mqa}. For simplicity these effects are not considered in this analysis.   

To test the agreement between the two spectra we multiply the M31 data by a scaling factor.  
This factor is then the ratio $J_r$.
Since the four GC background models yield significantly different spectra, we fit the scaling factor independently for each of them.

The best-fit scaling factor is determined using a $\chi^2$ fit. We account for upper limits (ULs) in the data by including an error function in the $\chi^2$ definition ~\cite{Lyu__2016,1986ApJ...306..490I,Karwin:2020tjw}
\bea
 \chi^2=
 \sum_{i=1}^m w_i^2
-2 \sum_{i=m+1}^{20}\mathrm{ln} \frac{1+\mathrm{erf}(w_i / \sqrt{2})}{2}
\label{chisqdef}
\eea
where 
\bea
w_i = {(\frac{y_i}{z_i} - J)\over \sigma_i^r}
\label{wi}
\eea
and 
\bea
\mathrm{erf}(z) =  \tfrac{2}{\sqrt{\pi}} \int_0^z e^{-t^{2}} \, \mathrm{d}t
\label{erf}
\eea

The first term on the right-hand side of Eq. (\ref{chisqdef}) is the classic definition of $\chi^2$, and the second term introduces the error function to quantify the fitting of ULs. The number of good data points is given by $m$, and the sum is over the 20 energy bins. 
Here $y_i$ is the flux from the GC and $z_i$ is the flux from M31, for the ith energy bin.

The error on the flux ratio $J_r$ is taken to
be
 \bea 
\sigma_i^r=\frac{\sigma_i^y}{z_i}
\eea
where we  use just the statistical 
error on the GC data, which we assume to be symmetric. This allows for a reasonable spectral comparison, and is further justified by the fact the uncertainty in the GC excess is dominated by the systematics. We note that in general a more sophisticated treatment of the errors may be appropriate  (e.g. ~\cite{Lyu__2016,Karwin:2020tjw}). However, we have tested different prescriptions for handling the error and in all cases we find that the results are qualitatively consistent.

We minimize the $\chi^2$ with respect to $J$,  and identify the 
minimum 
with  the optimized rescaling factor. 
The third column of Table \ref{table: Ratios} shows the best-fit results for each IEM. We note that the best-fit $J$-factor ratios are well within the bounds from section  \ref{J-facts}. There is a preference for smaller values of $J_r\sim 40$, corresponding to larger values of $J_{M31}$.  We will refer to these as the {\it model-independent} $J_r$ values (as these are found without reference to a specific DM annihilation model).

\begin{center}
\begin{table}[t]
\captionsetup{justification=centering}
\begin{tabular}{|c  c| c|} 
\hline
IEM    &
$J_r$  \\
\hline\hline
Pulsars, intensity-scaled
 & $48.96 \pm 2.31$    
\\
Pulsars, index-scaled
  & $36.63 \pm 3.37$     
\\
OB Stars, intensity-scaled  
    & $48.76 \pm 2.22$       
\\
OB Stars, index-scaled
 & $41.24 \pm 5.75$     
\\
\hline 
\end{tabular}
\caption{$J$-factor ratio for each IEM.
}
\label{table: Ratios}
\end{table}
\end{center}

\begin{figure}
    \includegraphics[width=\linewidth]{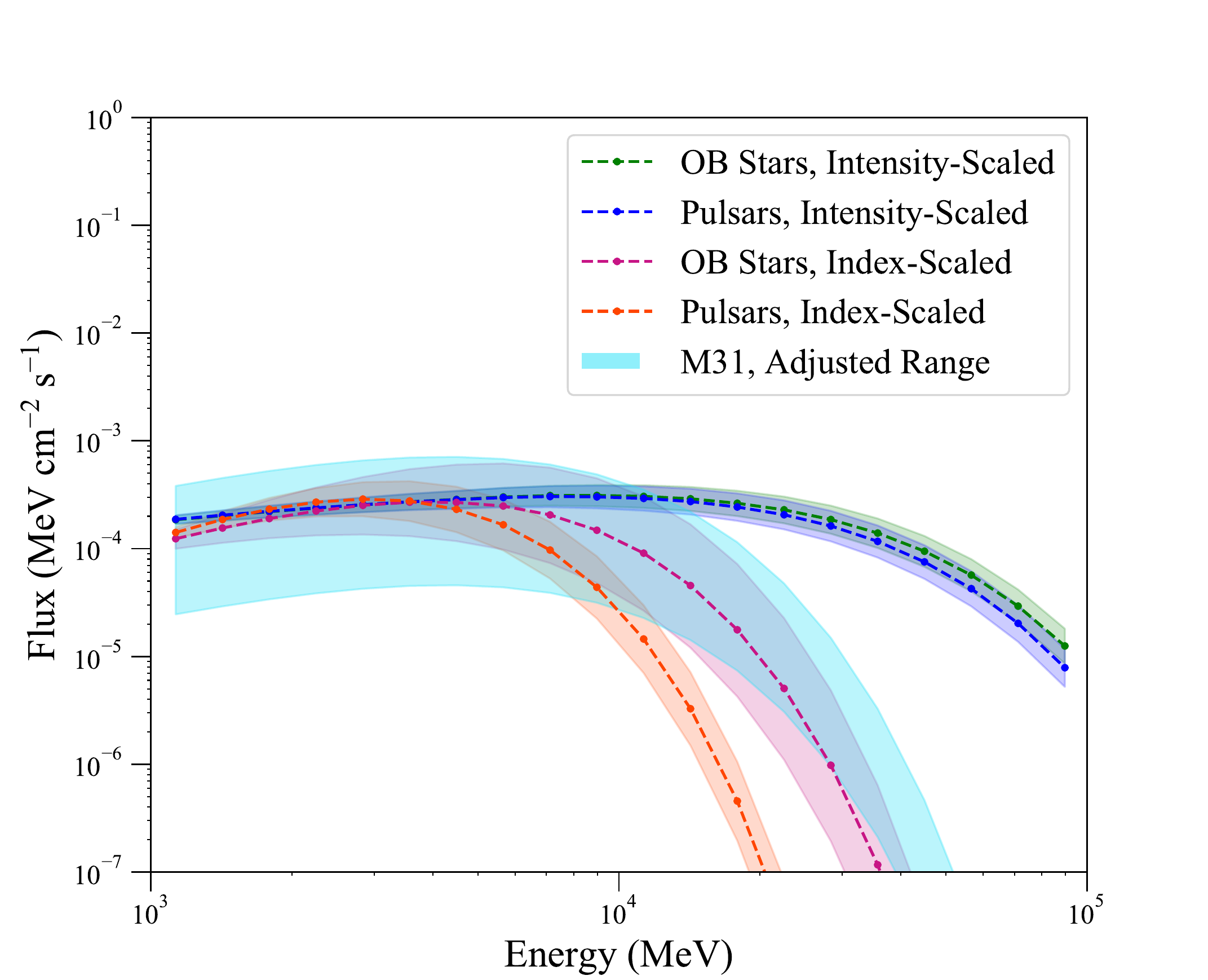}
    \caption{The blue band shows the range of M31 flux values scaled up by a $J$-factor ratio between 7.7 and 110. 
    Dashed lines show the four different GC IEMs with one sigma error bands.}
    \label{fig:M31_vs_all_noadjusted}
\end{figure}

\subsection{Spectral Comparisons}

We further examine the agreement between the M31 spectrum and the GC excess by scaling the M31 data by the 
best-fit $J$-ratio found above, and  comparing the two spectral shapes. 
These comparisons are shown in Fig.~\ref{fig:intensityvsM31}.
The top panel shows the rescaled M31 data compared to the GC excess for the index-scaled IEMs.  As can be seen, the spectra show excellent agreement. 

The bottom panel shows the intensity-scaled IEMs. 
As can be seen, there is a strong tension between the GC and M31 spectra at high energy (above $\sim$10 GeV). This is due to the existence of the so-called "high-energy tail" in the intensity-scaled IEMs. 
The nature of the high-energy tail of the GC excess has been investigated in numerous studies (e.g.~\cite{TheFermi-LAT:2015kwa,Calore:2014xka, Horiuchi:2016zwu,Linden:2016rcf}). It remains uncertain whether this feature is a true property of the signal or if it is due to mis-modeling of the background. When comparing the GC excess to the M31 excess, it is important to note that the two signals are extracted from very different regions of the galaxy, and thus they may not be directly comparable. In particular, this is the case when considering secondary emission, which depends on the astrophysical backgrounds. With that said, the M31 data does not possess a high-energy tail, and so seems to be in strong tension with those models. Indeed, this would be in general agreement with previous studies which have found that the high-energy tail is not very compatible with having a pure DM explanation~\cite{ Horiuchi:2016zwu,Linden:2016rcf}.

One can also examine whether a different choice of $J$-factor could ameliorate the  tension at high energies between the M31 excess and the intensity scaled GC excesses. To examine this, we find the range of
possibilities for the M31 flux, by rescaling it by the
maximum and minimum $J$-factors 
allowed from Ref.~\cite{Karwin:2020tjw}. 
Figure \ref{fig:M31_vs_all_noadjusted} shows the scaled M31 data compared to the GC excess for the four IEMs. As can be seen, the M31 data shows good agreement with the index-scaled IEMs, whereas there is still tension with the intensity-scaled IEMs.

\section{Dark Matter Models}
\label{sec:dark_matter_models}

In this section we perform a DM fit simultaneously to both signals. 
We will take a model where  DM is a real scalar field $\chi$ of mass $m_\chi$,
and consider various possibilities
for the dominant annihilation
process; specifically, we will
 consider
both two-body and four-body final states.
For the standard WIMP models the DM spectra were generated using PPPC ~\cite{pppc}. For the four-body annihilations the spectra were produced using FeynRules~\cite{feynrules}  and MadGraph~\cite{madgraph}, and showered with Pythia 8~\cite{pythia}.  The photons were binned in 20 logarithmically-spaced bins from 1$-$100 GeV, just as for the GC and M31 data. 

The predicted $\gamma$-ray flux from DM annihilation is given by
\bea
 \left. E^2 \frac{d \Phi }{dE}\right|_{GC} = 
  {\cal N}_{GC}  (E^2 \frac{dn}{dE}) 
  \\
   \left. E^2 \frac{d \Phi }{dE}\right|_{M31} = 
 \frac{{\cal N}_{GC}}{J_r}  (E^2 \frac{dn}{dE}) 
 \eea
 Here
 \bea
   {\cal N}_{GC} =
   J_{GC} \frac{\langle\sigma v\rangle}{4 \pi \eta m_{\chi}^2} 
\eea
where $\langle\sigma v\rangle$ is the velocity averaged cross section, $\eta$ is 2 (4) for conjugate (non-self conjugate) DM, $m_{\chi}$ is the DM mass,
and dn/dE is the number of $\gamma$-ray photons per annihilation. 

We perform a $\chi^2$ fit as in Eqs.~\ref{chisqdef}-\ref{erf}. The main difference is the definition of $w_i$. This quantity is defined separately for the GC and M31. For the GC:
\bea
w_i = {y_i - NE^2{dn\over de}\over \sigma^y_i}
\eea
where
$\sigma^y_{i}$
 is the 1-sigma error on the  GC flux and $y_i$ is the best fit  value of the GC flux
for a given IEM. Similarly for M31 we have
\bea
x_i = {z_i - J^{-1}NE^2{dn\over de}\over \sigma^z_i}
\eea
where the error $\sigma^z_{i}$ is the 1-sigma error on the M31 flux and $z_i$ is the best fit value of the M31 flux.

Finally, we define the total chi-squared  as
\bea
\chi^2_{tot} = \chi_{GC}^2+\chi_{M31}^2
\eea
We marginalize over $N$ in order to minimize this quantity with respect to $J_\mathrm{r}$ and $m_\chi$.
This is done separately for each GC IEM. 

\begin{figure}[t]
\begin{centering}
    \includegraphics[width=\linewidth]{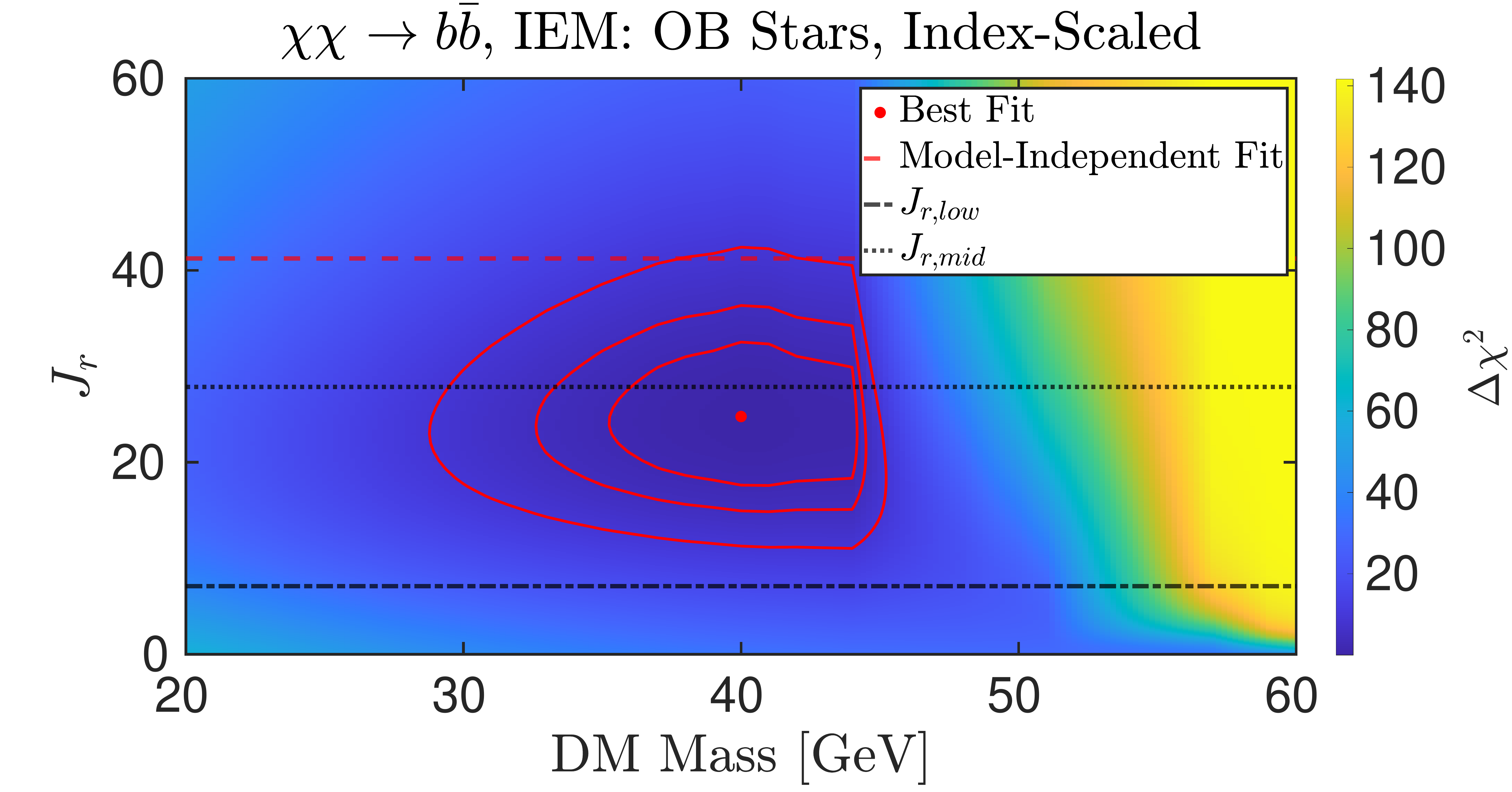}
    \caption{ $\Delta \chi^2$  for $\chi \chi \rightarrow b \bar{b}$ , for the  OB-Stars index-scaled IEM.  The red dot indicates
    the best fit point, and the contours are 1,2, and 3 $\sigma$ contours. 
    The dashed red line shows the  model independent 
    $J_r$ value.
Dash-dotted and dotted black lines show high and mean values of $J_r$
from section~\ref{J-facts}.}
    \label{fig:bb_obIndex_HG}
\end{centering}
\end{figure}

\begin{figure}
    \centering
  \includegraphics[width=\linewidth, height = 0.75\linewidth]{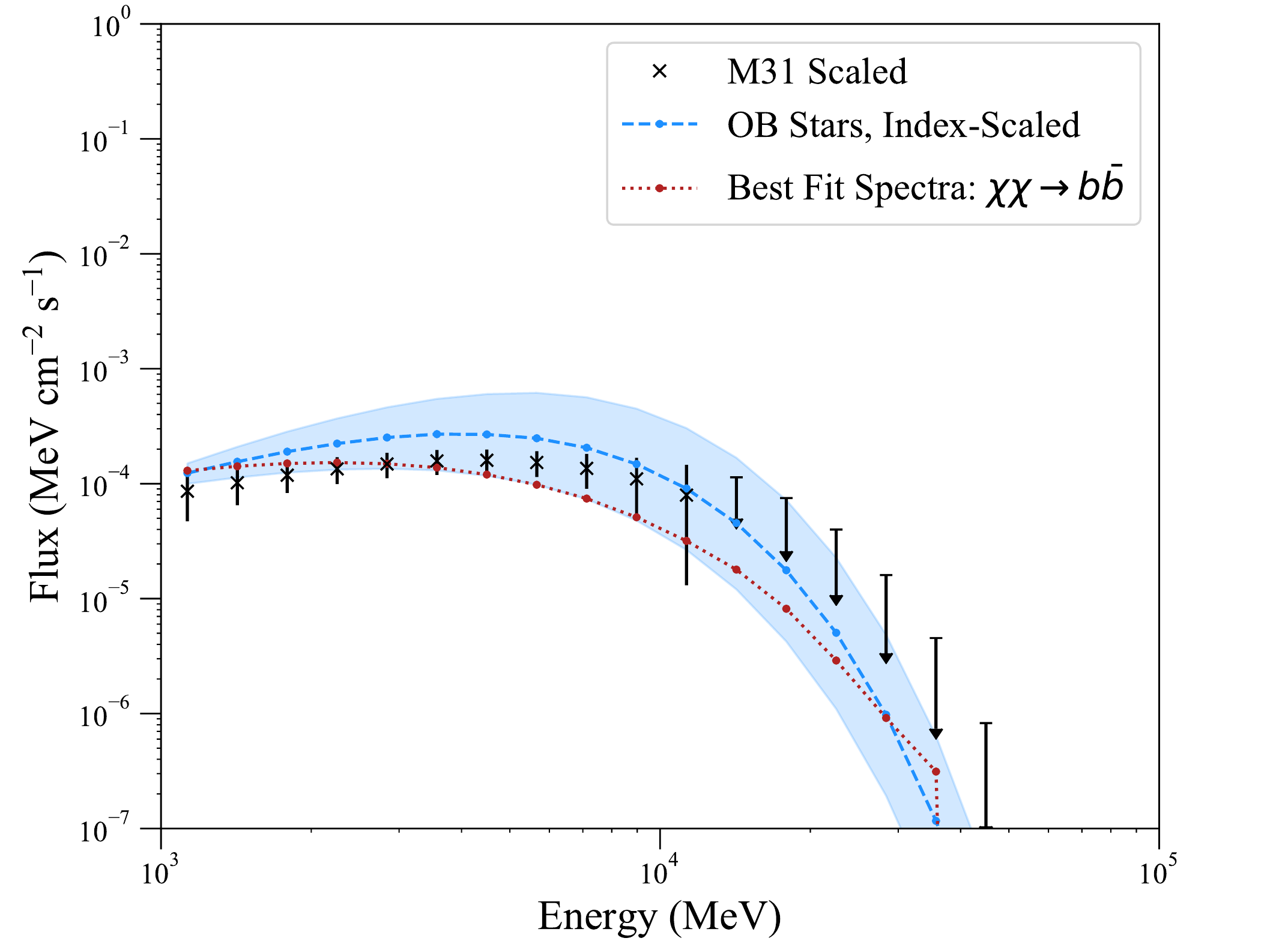}
  \caption{
Dashed lines show the GC excess with the  OB Stars, index IEM.
Black points show the M31 flux data scaled up by the appropriate ratio  $J_r$ taken from  Table~\ref{table: fits}. The dotted line shows the  corresponding best-fit model spectra
for $\chi \chi \rightarrow b \bar{b}$. 
}
\label{fig:intensityvsM31vsXX}
\end{figure}

Figure \ref{fig:bb_obIndex_HG} shows the results for the two-body annihilation to bottom quarks with the OB Stars, index-scaled IEM. The color scale indicates the value of $\Delta \chi^2$. The best-fit is shown with a red point, and also overlaid are the 1$\sigma$, 2$\sigma$, and 3$\sigma$ confidence contours, corresponding to $\Delta \chi^2 = 2.30, 4.61$ {and} $9.21$, respectively. For comparison, we also show the
model-dependent $J_r$ value from Table 1
and $J_{r, low},J_{r, mid}$ from  Section ~\ref{sec:data_selection}.
As can be seen, the $J_r$ corresponding to the DM fit is in good agreement with the range found in Ref.~\cite{2019ApJ...880...95K}. In Figure ~\ref{fig:intensityvsM31vsXX} we show the corresponding best-fit DM spectrum compared to the  GC and 
scaled M31 data.  

We have also extended our analysis to other possible annihilation modes; these results are presented in the Appendix. Specifically, we first considered other two-body annihilations where the DM annihilates to two tau leptons, and the case where the DM annihilates to
 two light quarks, which we take to be down quarks for concreteness. 
 Figure ~\ref{fig:intensityvsM31vsXX1} shows the  results for these
annihilation channels.

As mentioned above, direct detection and collider searches significantly constrain DM couplings. This has motivated the study of models where the DM is coupled to the SM quarks through a pseudoscalar mediator
\cite{karwin2017dark, escudero2017updated, Abdullah:2014lla, Rajaraman:2015xka}.  For example, one can consider a model
with a mediator $\phi$ 
and the interactions
\bea
{\cal L}_{int}=\chi^2\phi^2+\phi\overline{b}b
\eea
In this model, DM primarily annihilates to four b-quarks.
The precise annihilation mode depends on the coupling, for example if the mediator coupled as
 $\phi\overline{d}d$, there would be a annihilation to four d quarks. Generically we get a four-body annihilation.

Results for  some possible  4-body final states are shown in Figure ~\ref{fig:intensityvsM31vsXX2}. 
We note that the best fit DM  mass 
increases for the four body annihilation mode; this is expected because each quark has less energy.

The corresponding best-fit parameters for all models are summarized in Table ~\ref{table: fits}.

\section{Conclusion}

The GC excess, an excess 
of $\gamma$-ray photons from the GC, has been a long-standing potential signal of DM annihilation. However, the large 
astrophysical background and the potential existence of new sources 
makes it difficult to make definitive 
statements about the origin of this excess.
On the other hand, the M31 excess is from a region where astrophysical backgrounds (not associated with the conventional interstellar emission from 
the MW foreground) are not
expected to be large, and hence lends credence to the possibility  that the excess is indeed associated 
with DM annihilation, rather than an unknown astrophysical
background.

We have further examined  these two excesses, to see if their magnitudes and spectral shapes are consistent with DM annihilation.
The two signals are expected to be related by the ratio of the two $J$-factors.
The recent analysis of the M31 $J$-factor allows us to check this relation, and we
have found that indeed the excesses are
consistent with the determined $J$-factors.
The spectral shapes for the index-scaled IEMs are also in very good agreement. On the other hand, there is tension with the intensity-scaled IEMs due to the so-called high-energy tail. 

We also fit the excesses to a number
of DM models, where the DM annihilates to either  
two or four SM particles. We found that excellent fits
can be achieved both in two-body and four-body annihilations, as can be seen in the Appendix.

In summary, we have found that the M31 excess and the GC excess are mutually consistent with a dark matter origin. The DM models prefer a somewhat  higher value for the M31 $J$-factor, and prefer a particular IEM (the index-scaled models) for the GC. Currently, several DM models are consistent with the excesses. 

Future prospects to confirm the excess toward the outer halo of M31, and to better understand its nature, will crucially rely on improvements 
in modeling the interstellar emission towards M31. For the GC, the excess has been under investigation for many years now, and further improvements in the IEM will continue to play a significant role in better understanding the nature of the signal. Additionally,  working towards a better understanding of the possible point-like nature of the excess will be key. Improved sensitivity from other indirect detection constraints will also continue to play an important role in DM interpretations of the two signals, and likewise for constraints from direct detection.    Further analysis of these complementary signals  would be extremely interesting, and could shed light on the nature of DM.

\section*{Acknowledgements}
We are especially grateful to Simona Murgia for many detailed explanations of how to interpret and analyze the published data from \textit{Fermi}-LAT, and for valuable feedback on our analysis.
This work was supported in part by NSF Grant No.~PHY-1915005.

\bibliography{main6.bib}

\pagebreak

\appendix

\begin{figure*}
  \centering
  \textbf{Appendix A: Other Annihilation Channels}\par\medskip
  \includegraphics[width=.4\linewidth]{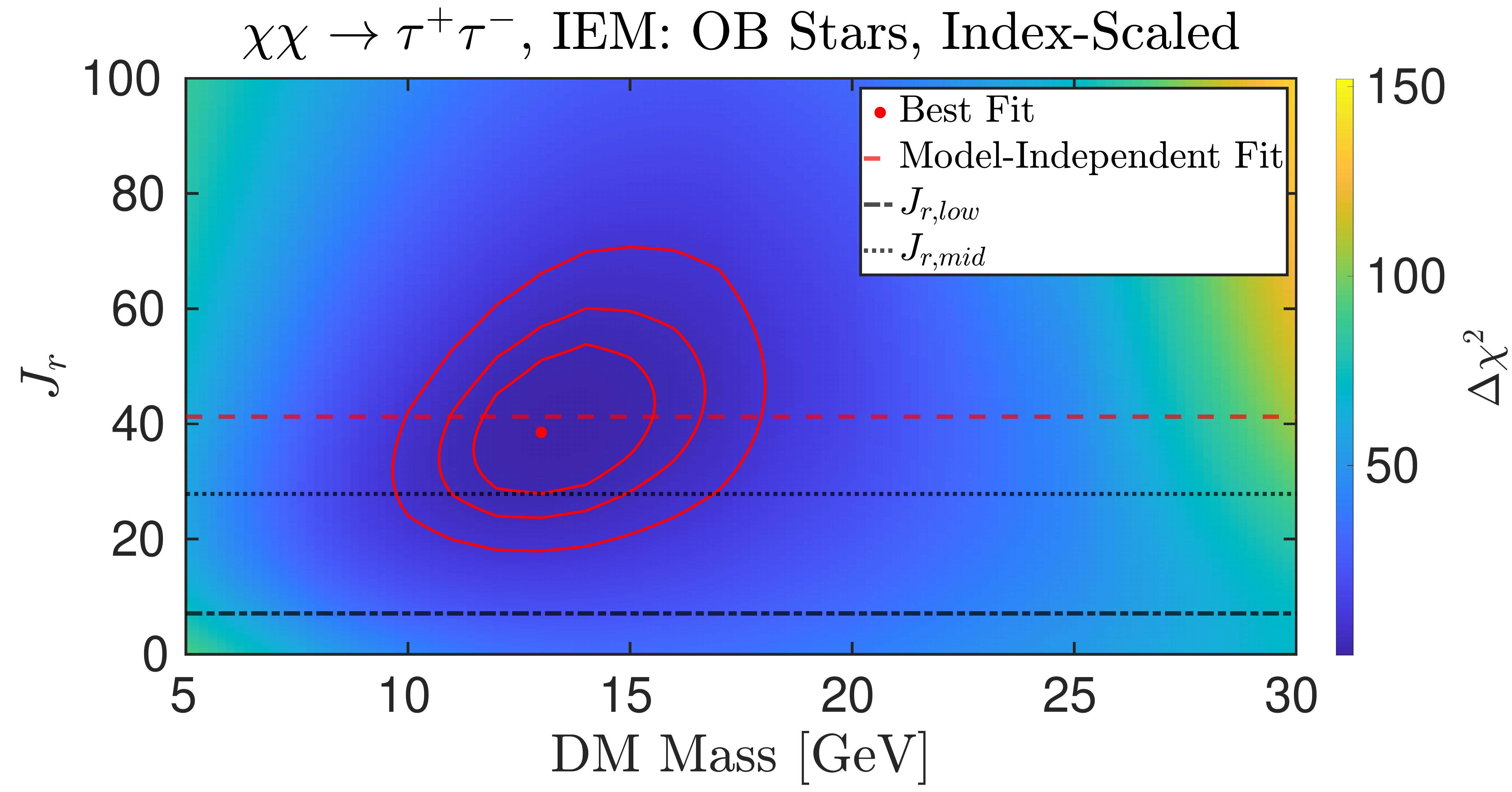}
    \includegraphics[width=.4\linewidth]{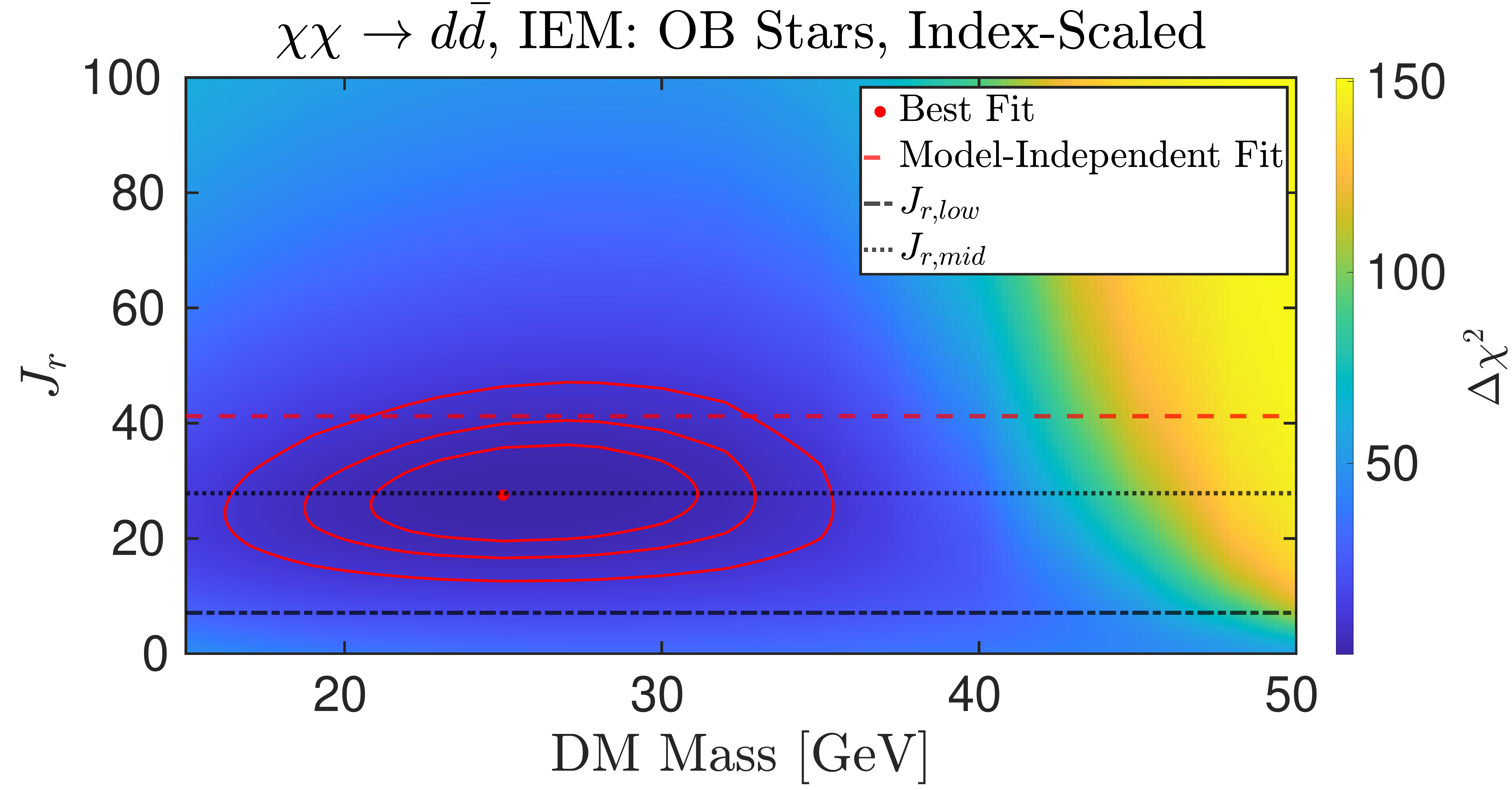}
  \centering
      \includegraphics[width=.4\linewidth]{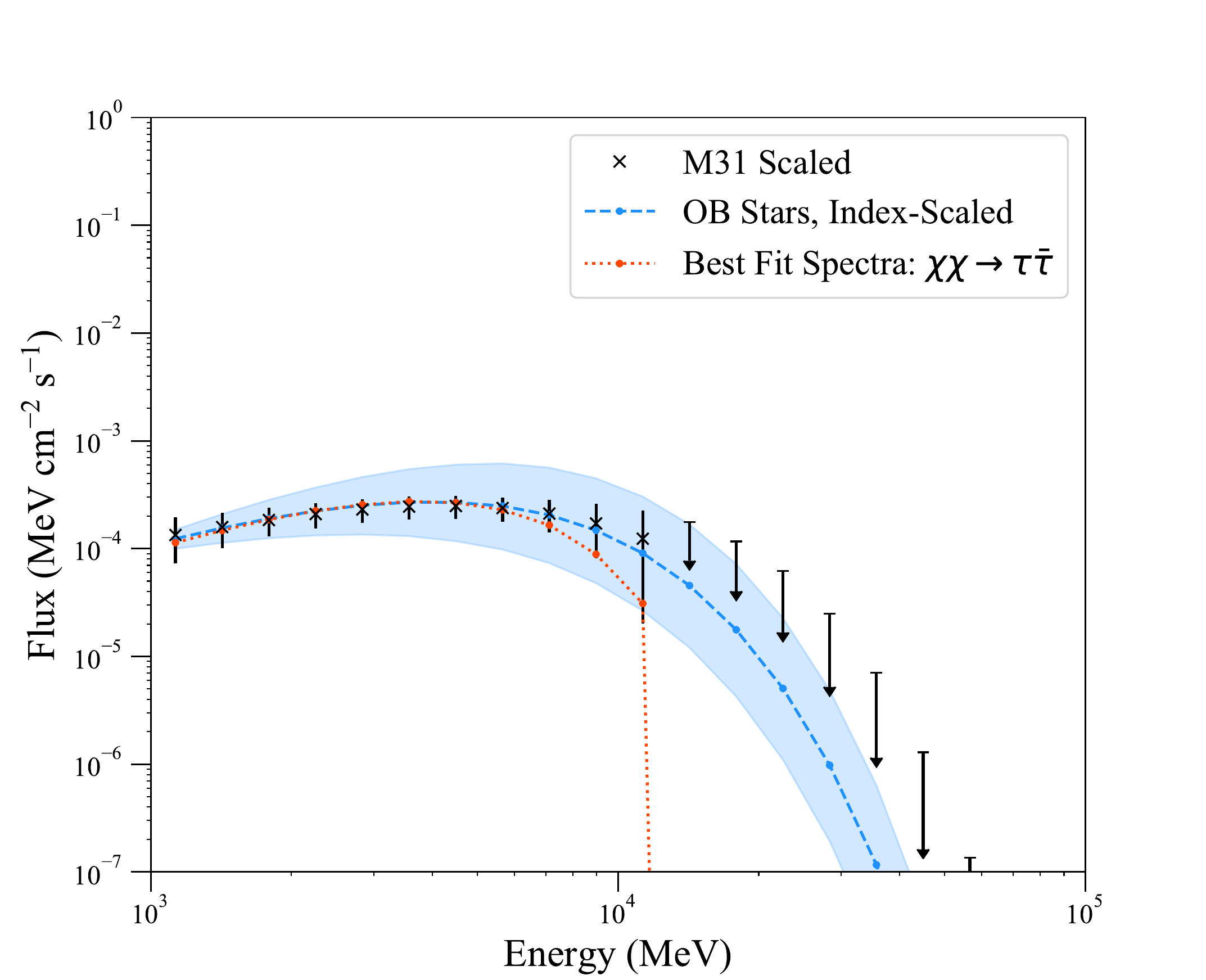}
  \includegraphics[width=.4\linewidth]{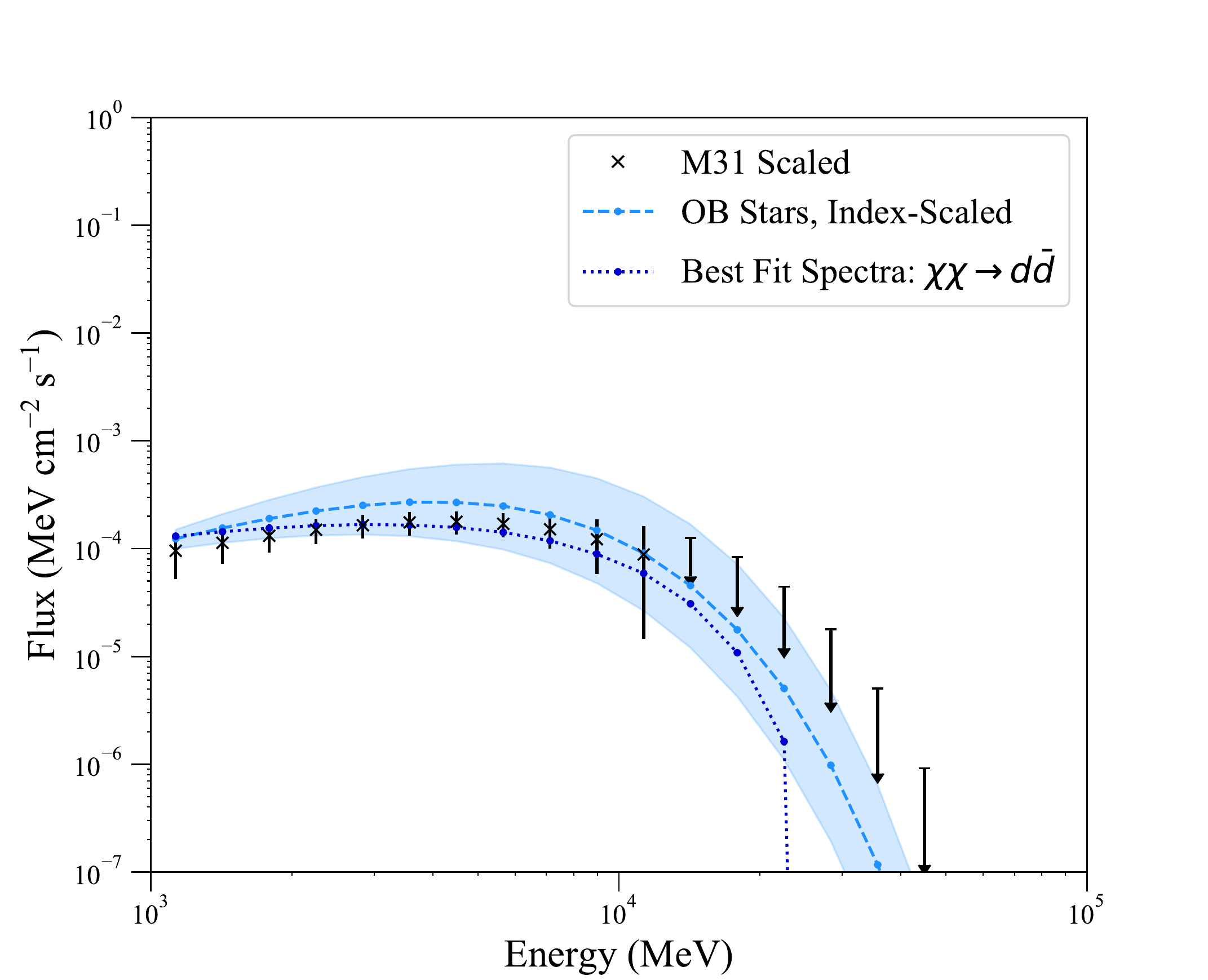}
  \caption{The left panels are similar to Figures \ref{fig:bb_obIndex_HG} and 
  \ref{fig:intensityvsM31vsXX} for $\chi \chi \rightarrow \tau \bar{\tau}$, and the right panels are similar to Figures \ref {fig:bb_obIndex_HG} and \ref{fig:intensityvsM31vsXX}
  for $\chi \chi \rightarrow d \bar{d}$}
\label{fig:intensityvsM31vsXX1}
\end{figure*}

\begin{figure*}
  \centering
  \includegraphics[width=.4\linewidth]{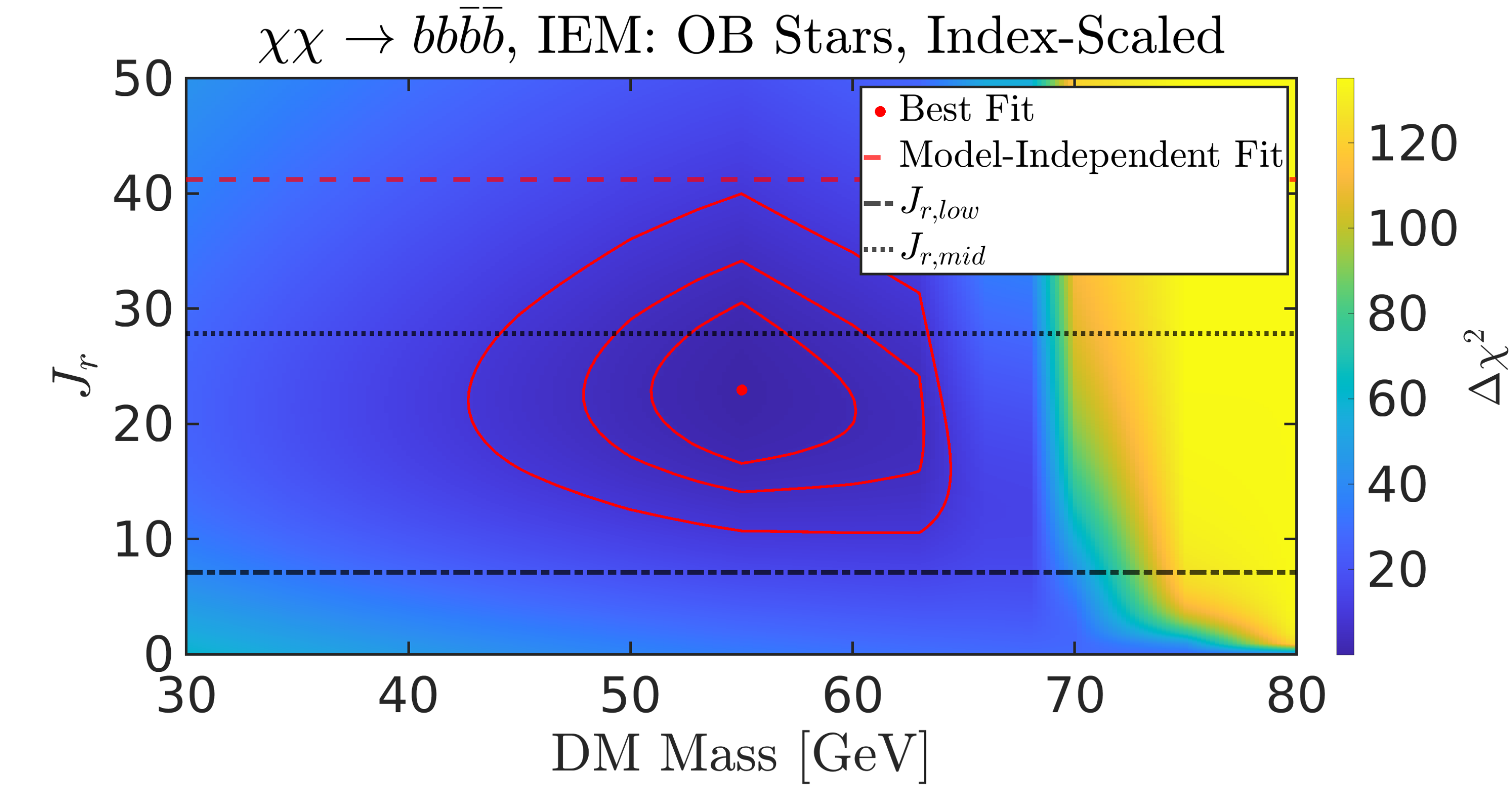}
    \includegraphics[width=.4\linewidth]{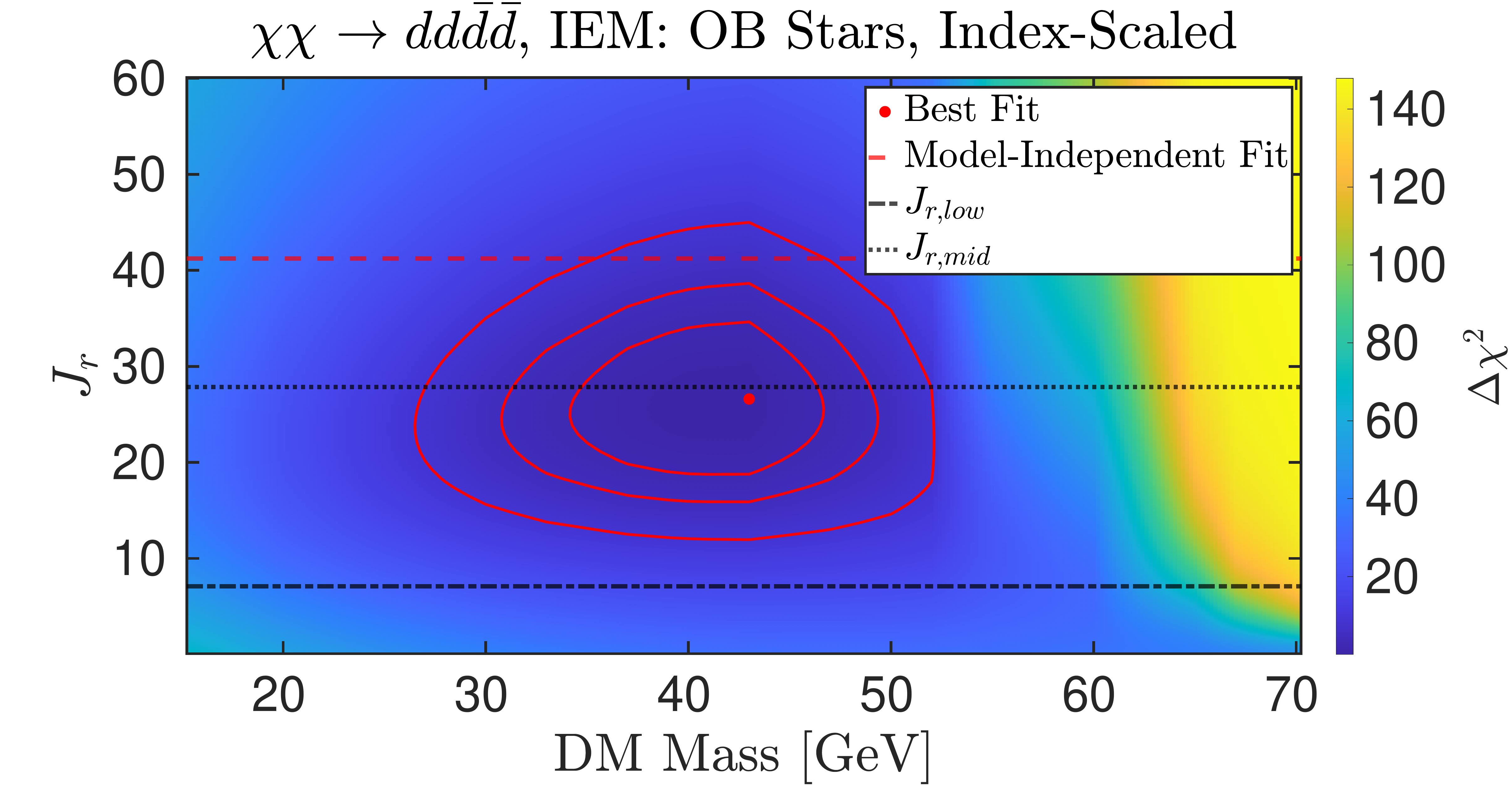}
  \centering
      \includegraphics[width=.4\linewidth]{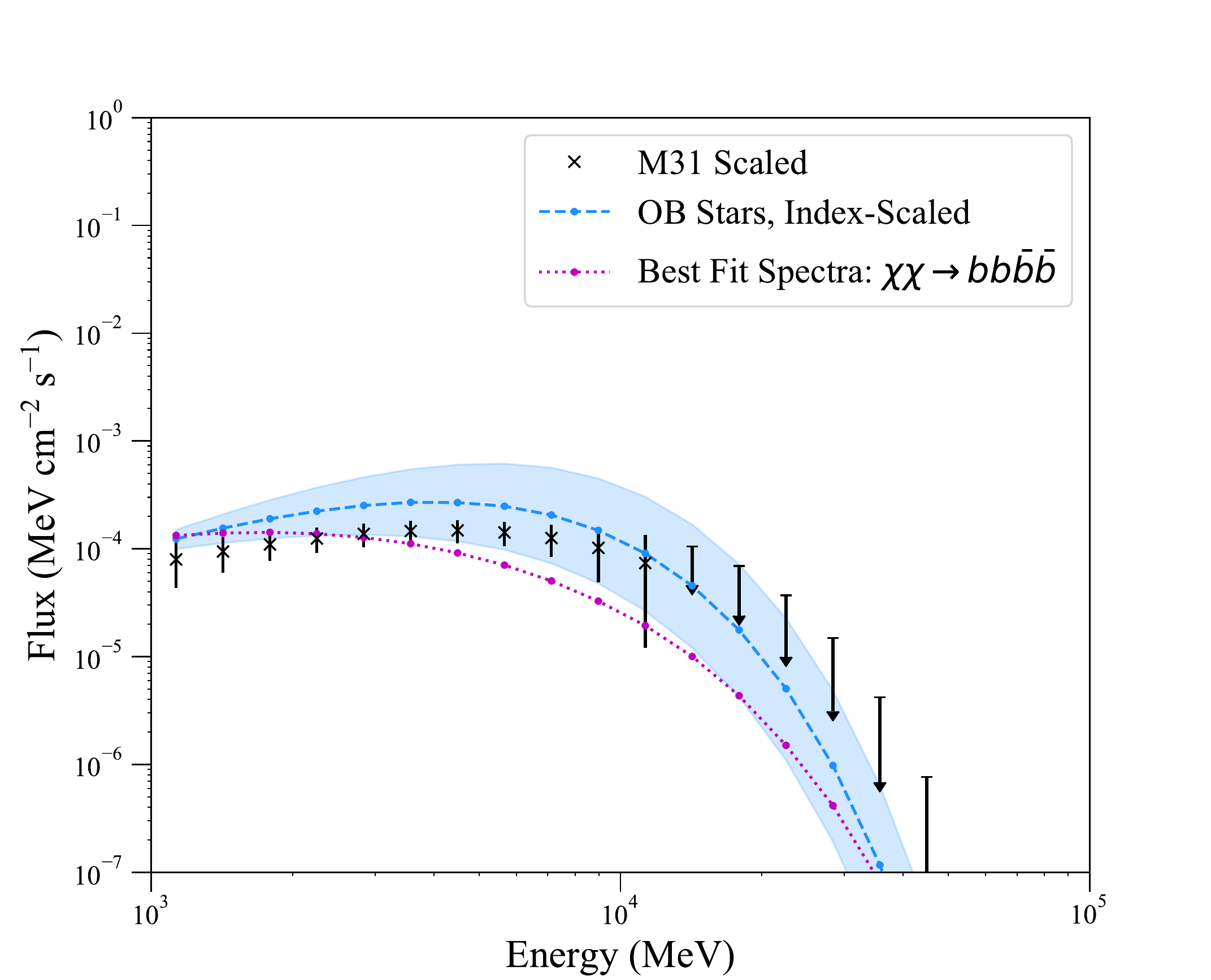}
  \includegraphics[width=.4\linewidth]{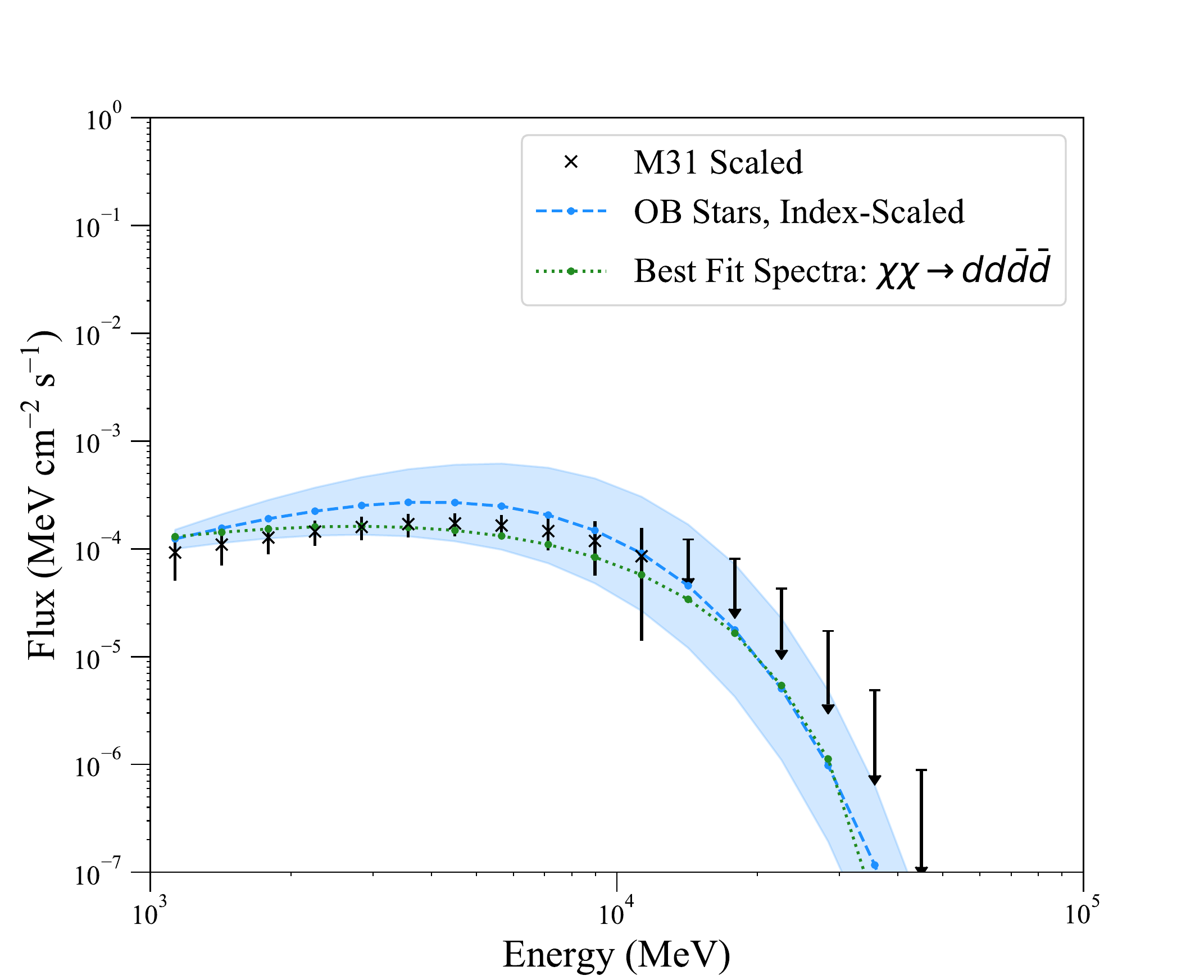}
  \caption{
Similar to  Fig ~\ref{fig:intensityvsM31vsXX1}, but for $\chi \chi \rightarrow bb \bar{b}\bar{b}$ and $\chi \chi \rightarrow dd \bar{d}\bar{d}$ respectively.
}
\label{fig:intensityvsM31vsXX2}
\end{figure*}

\begin{table*}[t]
\captionsetup{justification=centering}
\begin{center}
\begin{tabular}{ |c c c c c c|} 
\hline
DM Model & IEM & $m_{\chi}$ [GeV]  & ${\cal{N}}_{GC}\times 10^{8}$ [cm$^{-2}$ s$^{-1}$]  & $J_r$ & $\chi^2_{red}$ \\
\hline
\multirow{4}{4em}{$b\bar{b}$} & Pulsars, intensity-scaled & 
$57_{-2.1}^{1.3}$ 
& $2.6_{-0.14}^{+0.14}$ & $45.9_{-6.9}^{+8.1}$  & 2.00  \\[2ex] 
& Pulsars, index-scaled & $22_{-0.9}^{+1.9}$  & $3.9_{-0.31}^{+0.31}$ & $25.7_{-4.8}^{+6.7}$  & 1.72 \\ [2ex] 
& OB Stars, intensity-scaled & $57_{-1.7}^{+1.3}$  & $2.7_{-0.16}^{+0.16}$ & $45.9_{-6.9}^{+8.0}$  & 2.4 \\ [2ex]
& OB Stars, index-scaled & $40_{-4.7}^{+4.1}$  & $2.1_{-0.10}^{+0.10}$ & $24.8_{-7.2}^{+7.8}$  & 1.01 \\[2ex]
\hline
\multirow{4}{4em}{$d\bar{d}$} & Pulsars, intensity-scaled & $43_{-4.9}^{+3.7}$  & $3.6_{-0.16}^{+0.16}$ & $51.4_{-7.2}^{+9.8}$  & 1.43 \\ [2ex]
& Pulsars, index-scaled & $17_{-1.3}^{+1.3}$  & $5.0_{-0.24}^{+0.24}$ & $29.4_{-5.4}^{+6.5}$  & 0.99 \\ [2ex]
& OB Stars, intensity-scaled & $45_{-5.5}^{+3.1}$  & $3.6_{-0.17}^{+0.17}$ & $53.2_{-7.7}^{+9.9}$  & 1.72 \\ [2ex]
& OB Stars, index-scaled & $25_{-4.1}^{+6.1}$  & $3.3_{-0.01}^{+0.01}$ & $27.5_{-8.0}^{+8.3}$  & 0.81 \\[2ex]
\hline
\multirow{4}{4em}{$bb\bar{b}\bar{b}$} & Pulsars, intensity-scaled & $81_{-3.0}^{+0.3} $ & $1.76_{-0.11}^{+0.11}$ & $42.2_{-5.8}^{8.4}$ & 2.40 \\ [2ex]
& Pulsars, index-scaled & $36_{-3.5}^{+1.1}$  & $2.4_{-0.19}^{+0.19}$ & $24.8_{-5.1}^{+5.9}$  & 1.92 \\ [2ex]
& OB Stars, intensity-scaled & $81_{-2.5}^{+0.3}$  & $1.8_{-0.12}^{+0.12}$ & $42.2_{-5.7}^{+8.4}$  & 2.85 \\ [2ex]
& OB Stars, index-scaled & $55_{-4.1}^{+4.9}$ & $1.4_{0.01}^{+0.01}$ & $22.9_{-6.3}^{+7.6}$  & 1.19 \\[2ex]
\hline
\multirow{4}{4em}{$dd\bar{d}\bar{d}$} & Pulsars, intensity-scaled & $67_{-6.7}^{+3.9}$ & $2.2_{-0.1}^{+0.1}$ & $50.5_{-7.5}^{+9.1}$ & 1.57\\ [2ex]
& Pulsars, index-scaled & $25_{-2.3}^{+1.3}$ & $3.1_{-0.18}^{+0.18}$ & $27.5_{-5.0}^{+6.6}$ & 1.24\\ [2ex]
& OB Stars, intensity-scaled & $67_{-5.1}^{+4.3}$& $2.2_{-0.11}^{+0.11}$ & $50.5_{-7.5}^{+9.0}$ & 1.89 \\ [2ex]
& OB Stars, index-scaled & $43_{-8.7}^{+3.7}$ & $1.9_{-0.01}^{+0.01}$ & $26.6_{-7.8}^{+8.0}$  & 0.87 \\[2ex]
\hline
\multirow{4}{4em}{$\tau^+\tau^-$} & Pulsars, intensity-scaled & $15_{-1.7}^{+1.1}$  & $15.7_{-1.03}^{+1.03}$ & $52.3_{-7.3}^{+9.7}$  & 2.51  \\ [2ex]
& Pulsars, index-scaled & $12_{-1.5}^{+0.5}$  & $15.7_{-0.53}^{+0.53}$ & $43.1_{-7.7}^{+9.3}$ & 0.63  \\ [2ex]
& OB Stars, intensity-scaled & $15_{-1.1}^{+1.5}$ & $15.6_{-1.1}^{+1.1}$ & $52.3_{-7.7}^{+9.1}$ & 2.90 \\ [2ex]
& OB Stars, index-scaled & $13_{-1.5}^{+2.3}$ & $15.5_{-1.14}^{+1.14}$ & $38.5_{-10.6}^{+12.6}$  & 0.78 \\[2ex]
\hline
\end{tabular}
\caption{
Best-fits for $m_{\chi}$, ${\cal N}_{GC}$ , and the ratio $J_r$ for various annihilation channels.
\label{table: fits}}
\end{center}
\end{table*}

\end{document}